\documentclass[twocolumn,showpacs,preprintnumbers,amsmath,amssymb]{revtex4}

\usepackage{graphicx}
\usepackage{dcolumn}
\usepackage{bm}

\newcommand{\kp}{\mathbf{k}_\perp}
\newcommand{\p}{\perp}
\newcommand{\ssh}{\!\!\!/}

\begin{document}

\title{Non-zero transversity distribution of the pion in a quark-spectator-antiquark model}
\author{Zhun L\"{u}}
\affiliation{Department of Physics, Peking University, Beijing
100871, China}

\author{Bo-Qiang Ma}\email[Corresponding author. Electronic address: ]{mabq@phy.pku.edu.cn}
\affiliation{ CCAST(World Laboratory), P.O. Box 8730, Beijing 100080, China\\
and Department of Physics, Peking University, Beijing 100871,
China}

\begin{abstract}
We calculate the non-zero (na\"{i}ve) T-odd transverse momentum
dependent transversity distribution $h_1^{\perp}(x,\kp^2)$ of the
pion in a quark-spectator-antiquark model. The final-state
interaction is modelled by the approximation of one gluon exchange
between the quark and the antiquark spectator. Using our model
result we estimate the unsuppressed cos2$\phi$ azimuthal asymmetry
in unpolarized $\pi^-p$ Drell-Yan process. We find that the
transverse momentum dependence of $h_1^{\perp}(x,\kp^2)$ of the
pion is the same as that of $h_1^{\perp}(x,\kp^2)$ of the proton
calculated from the quark-scalar-diquark model, although the $x$
dependencies of them are different from each other. This suggests
a connection between cos2$\phi$ asymmetries in Drell-Yan processes
with different initial hadrons.
\end{abstract}


\maketitle

\section{Introduction}
Recently the study of transverse momentum dependent distribution
functions is among the special issues in hadronic physics. Of
particular interest, are two leading-twist (na\"{i}ve)
time-reversal odd transverse momentum dependent distribution
functions: Sivers function~\cite{sivers90,abm95}
$f_{1T}^{\perp}(x,\kp^2)$ and its chiral-odd partner
$h_1^{\perp}(x,\kp^2)$~\cite{bm98,boer99}. Sivers function
represents the unpolarized parton distribution in a transversely
polarized hadron, while $h_1^{\perp}(x,\kp^2)$ denotes the parton
transversity distribution in an unpolarized hadron. One main
motivation to investigate these two distributions is that they are
the possible sources of the unsuppressed azimuthal asymmetries
observed in hadronic reactions. The former distribution function
was proposed first by Sivers~\cite{sivers90} to illustrate that it
can lead to large single-spin azimuthal asymmetries. This
nontrivial correlation between the transverse momentum of the
quark and the polarization of the hadron was thought to be
forbidden by time-reversal invariance~\cite{collins93}. Recently a
direct calculation~\cite{bhs02a,bhs02b} of Sivers asymmetry by
inclusion of final-state (in semi-inclusive deep inelastic
scattering(SIDIS)) or initial-state interaction (in Drell-Yan
process) shows that the asymmetry is in principle non-zero. Then
it was found that the presence of the Wilson lines in the
operators defining the parton densities allows for the Sivers
effect without a violation of time-reversal
invariance~\cite{collins02}, and the final- or initial-state
interaction can be factorized into a full gauge-invariance
definition of transverse momentum dependent distribution
functions~\cite{jy02,bmp03}.

 These theoretical developments open a
wide range of phenomenological applications. Several model
calculations~\cite{yuan03,gg03,bsy04,lm04} of Sivers function have
been performed to  estimate single-spin asymmetries in SIDIS
process, which is under investigation by current
experiment~\cite{hermes04}. On the other hand, it is
shown~\cite{gg03} that non-zero $h_1^{\perp}(x,\kp^2)$ can arise
from the same mechanism which produces $f_{1T}^{\perp}(x,\kp^2)$.
It has been demonstrated~\cite{boer99} that $h_1^{\perp}(x,\kp^2)$
can account for the substantial cos$2\phi$ asymmetries in
unpolarized Drell-Yan lepton pair production from pion-nucleon
scattering: $\pi^-N\rightarrow \mu^+\mu^-X$~\cite{na10,conway89}.
In Ref.~\cite{bbh03}, Boer, Brodsky, and Hwang  computed
$h_1^{\perp}(x,\kp^2)$ of the proton in a quark-scalar-diquark
model within soft gluon exchange. They found that
$h_1^{\perp}(x,\kp^2)$ is equal to $f_{1T}^{\perp}(x,\kp^2)$
obtained from the same model. Then the maximum magnitude of the
cos$2\phi$ asymmetries in $p\bar{p}\rightarrow l\bar{l}X$ is
estimated to be $\sim$30$\%$, by using the calculated
$h_1^{\perp}(x,\kp^2)$. In this paper, we perform the first
computation on $h_1^{\perp}(x,\kp^2)$ of the pion (denoted as
$h_{1\pi}^{\perp}(x,\kp^2)$) in a quark-spectator-antiquark model
in presence of final-state interaction, in similar to the
quark-scalar-diquark model of the proton. We find that the
transverse momentum dependence of $h_{1\pi}^{\perp}(x,\kp^2)$ in
our model is the same as that of $h_1^{\perp}(x,\kp^2)$ of the
proton from the quark-scalar-diquark model, although the $x$
dependence is different. This feature allows one to expect that
$h_{1\pi}^{\perp}(x,\kp^2)$ and $h_1^{\perp}(x,\kp^2)$ are closely
related. With the present model result we investigate the
cos2$\phi$ asymmetry in unpolarized $\pi^-p$ Drell-Yan process and
obtain an unsuppressed result. The shape of the asymmetry is
similar to the cos$2\phi$ asymmetries in $p\bar{p}\rightarrow
l\bar{l}X$, estimated in Ref.~\cite{bbh03}. The result suggests a
connection between cos2$\phi$ asymmetries in Drell-Yan processes
with different initial hadrons.

\section{Calculation of $h_{1\pi}^{\perp}(x,\kp^2)$ of the pion}
In this section, we present the calculation $h_1^{\perp}(x,\kp^2)$
of the pion. We start our computation from the quark light-cone
correlation function of the pion in Feynman gauge (we will perform
our calculation in this gauge):
\begin{eqnarray}
\Phi_{\alpha\beta}(x,\kp)&=&\int \frac{d \xi^- d^2
\mathbf{\xi}_\perp}{(2\pi)^3}e^{ik\cdot\xi}\langle
P_\pi|\bar{\psi}_\beta(0)
\mathcal{L}_0^\dag(0^-,\infty^-)\nonumber\\
&&\times\mathcal{L}_\xi(\infty^-,\xi^-)\psi_\alpha(\xi)|P_\pi\rangle\bigg{|}_{\xi^+=0}.
\end{eqnarray}
We use notation $a^\pm=a^0\pm a^3$, $a\cdot
b=\frac{1}{2}(a^+b^-+a^-b^+)-\mathbf{a}_\perp\cdot\mathbf{b}_\perp$.
The pion momentum is denoted by $P_\pi^{\mu}=(P_\pi^+, P_\pi^-,
\mathbf{P}_{\pi\perp})=(P_\pi^+, M/P_\pi^+, \mathbf{0}_{\perp})$.
$\mathcal{L}_0(0,\infty)$ is the path-ordered exponential (Wilson
line) with the form:
\begin{equation}
\mathcal{L}_0(0,\infty)=\mathcal{P}~\textmd{exp}\left
(-ig\int_{0^-}^{\infty^-}d\xi^-\cdot
A(0,\xi^-,\mathbf{0}_\perp)\right).
\end{equation}
Releasing the constraint of (na\"{i}ve) time-reversal invariance
and keeping parity invariance and hermiticity, the quark
correlation function of the pion can be parameterized into a set
of transverse momentum dependent distribution functions in leading
twist as follows~\cite{tm96,bm98}
\begin{equation}
\Phi(x,\kp)=\frac{1}{2}\left
[f_{1\pi}(x,\kp^2)n\ssh+h_{1\pi}^\perp(x,\kp^2)\frac{\sigma_{\mu\nu}\kp^\mu
n^{\nu}}{M_\pi}\right ],
\end{equation}
where $n$ is the light-like vector with components $(n^+, n^-,
\mathbf{n}_\perp)=(1, 0, \mathbf{0}_\perp)$,
$\sigma_{\mu\nu}=\frac{i}{2}[\gamma^\mu,\gamma^\nu]$,
$f_{1\pi}(x,\kp^2)$ and $h_{1\pi}^{\perp}(x,\kp^2)$ denote the
unpolarized quark distribution and the quark transversity of the
pion, respectively. Knowing $\Phi_\pi(x,\kp)$, one can obtain
these distributions from equations:
\begin{eqnarray}
&f_{1\pi}(x,\kp^2)&=\textmd{Tr}[\Phi(x,\kp)\gamma^+];\\
&\frac{2h_{1\pi}^{\perp}(x,\kp^2)\mathbf{k}_\p^i}{M_\pi}&=
\textmd{Tr}[\Phi(x,\kp)\sigma^{i+}].
\end{eqnarray}

\begin{figure}
\begin{center}
\scalebox{0.85}{\includegraphics{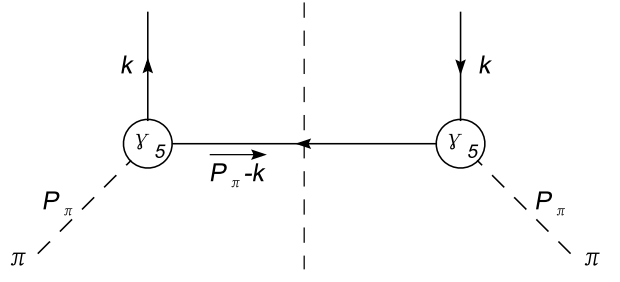}}\caption{\small Diagram
which gives $\Phi$ of the
 pion in the quark-spectator-antiquark model in lowest order.}\label{lo}
\end{center}
\end{figure}

We will calculate above distribution functions in the
quark-spectator-antiquark model. It is similar to the
quark-scalar-diquark model for calculating Sivers function and
$h_1^{\p}$ of the proton, and the differences are that the
intermediate state here is the constituent antiquark instead, as
shown in Fig.~1, and that the pion-quark-antiquark interaction is
modelled by pseudoscalar coupling:
\begin{equation}
\mathcal{L}_I=-g_\pi\bar{\psi}\gamma_5\psi\varphi-e_2\bar{\psi}\gamma^u\psi
A_\mu,
\end{equation}
in which $g_\pi$ is the pion-quark-antiquark coupling constant,
and $e_2$ is the charge of the antiquark. $f_{1\pi}(x,\kp^2)$ can
be calculated from the lowest order $\Phi(x,\kp)$ without the
path-ordered exponential. From Fig.~\ref{lo} we obtain
\begin{eqnarray}
f_{1\pi}(x,\kp^2)&=&-\frac{1}{4(1-x)P_\pi^+}\frac{g_\pi^2}{2(2\pi)^3}\sum_s\bar{v}^s\gamma_5
\frac{k\ssh+m}{k^2-m^2}\nonumber\\
&&\times\gamma^+\frac{k\ssh+m}{{k^2-m^2}}\gamma_5v^s,
\end{eqnarray}
where $m$ is the mass of the outgoing quark which is the same of
the antiquark, $v^s$ is the spinor of the antiquark and the quark
momentum $k=(xP^+,(\kp^2+m^2)/xP^+,\kp)$. We take the spin sum as
$\sum_s v^s\bar{v}^s=(P\ssh_\pi-k\ssh-m)$, which is a little
different from the spin sum adopted in Ref.~\cite{jmr97}.
Immediately we arrive at
\begin{eqnarray}
f_{1\pi}(x,\kp^2)&=&\frac{g_\pi^2}{2(2\pi)^3}\frac{(1-x)[\kp^2+(1+x)^2m^2]}
{(\kp^2+m^2-x(1-x)M_\pi)^2}\nonumber\\
&=&C_{\pi}\frac{\kp^2+D_{\pi}}{(\kp^2+B_{\pi})^2},
\end{eqnarray}
where we set $C_{\pi}=g_{\pi}^2(1-x)/[2(2\pi)^3]$,
$D_{\pi}=(1+x)^2m^2$ and
\begin{equation}
B_\pi=m^2-x(1-x)M_\pi^2.\label{bpi}
\end{equation}

\begin{figure}
\begin{center}
\scalebox{0.85}{\includegraphics{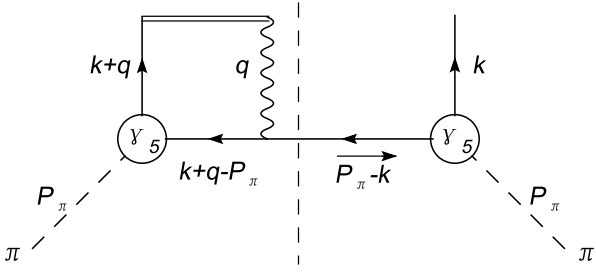}}\caption{\small Diagram
which yields $\Phi$ with final-state interaction modelled by one
gluon exchange.}\label{fsi}
\end{center}
\end{figure}

 The T-odd distribution $h_{1\pi}^\p(x,\kp^2)$,
however, is absent in the lowest order $\Phi(x,\kp)$. In order to
produce this T-odd distribution, the path-ordered exponential
which ensures gauge invariance of the distribution function has to
be included. The exponential serves as the final-state
interaction~(FSI) or initial-state interaction~(ISI) between the
struck quark and the remnant of the hadron, which is also viewed
as the soft gluon scattering, to provide nontrivial phase to
generate T-odd distribution function. In our calculation we expand
path-ordered exponential to first order, means that the final-
or/and initial-state interaction is modelled by one gluon
exchange, as shown in Fig.~\ref{fsi}. Thus the nonzero
$h_{1\pi}^\p(x,\kp^2)$ can be calculated from the expression
\begin{eqnarray}
&&\frac{2h_{1\pi}^\p(x,\kp^2)\kp^i}{M_\pi}=\sum_{\bar{q}}\frac{1}{2}\int\frac{d\xi^-d\xi_\p}{(2\pi)^3}
e^{ik\cdot\xi}\langle P_\pi|\bar{\psi}_\beta(0)|\bar{q}\rangle\langle \bar{q}|\nonumber\\
&&\times\left (-ie_1 \int_{\xi^-}^{\infty^-}
A^+(0,\xi^-,\mathbf{0}_\p)d\xi^-\right )
\sigma^{i+}_{\beta\alpha}\psi_{\alpha}(\xi)|P_\pi\rangle\bigg{|}_{\xi^+=0}\nonumber\\
&&+h.c.,
\end{eqnarray}
in which $|\bar{q}\rangle$ represents the antiquark spectator
state, and $e_1$ is the charge of the struck quark. In momentum
space we write down:
\begin{eqnarray}
&&\frac{2h_{1\pi}^\p(x,\kp^2)\kp^i}{M_\pi}=\frac{-ie_1e_2}{8(2\pi)^3(1-x)P_\pi^+}
\sum_{s}\int\frac{d^4q}{(2\pi)^4}\bar{v}^s\gamma_5\nonumber\\
&&\times\frac{k\ssh+m}{k^2-m^2}
\sigma^{i+}\frac{k\ssh+q\ssh+m}{(k+q)^2-m^2}\gamma_5
\frac{k\ssh+q\ssh-P\ssh_\pi+m}{(k+q+P_\pi)^2+m^2}\nonumber\\
&&\times\gamma^+v^s\frac{1}{q^++i\epsilon}\frac{1}{q^2-i\epsilon}+h.c.
. \label{h1t}
\end{eqnarray}
The $\gamma^+$ in the second line of Eq.~(\ref{h1t}) comes from
the antiquark-gluon interaction vertex. The $q^-$ integral can be
done from the contour method. $q^+$ integral is realized from
taking the imaginary part of the eikonal propagator:
$1/(q^++i\epsilon)\rightarrow -i\pi\delta(q^+)$. Then we obtain
\begin{eqnarray}
\frac{h_{1\pi}^\p(x,\kp^2)\kp^i}{M_\pi}&=&-\frac{g_\pi^2e_1e_2(1-x)}{2(2\pi)^3(\kp^2+B_\pi)}
\int\frac{d^2q_\p}{(2\pi)^2}\nonumber\\
&&\times\frac{2m\mathbf{q}_\p^i}{q_\p^2[(\kp+q_\p)^2+B_\pi]}.
\end{eqnarray}

To arrive at above equation, we have calculated the trace in the
numerator in Eq.~(\ref{h1t}) as follows
\begin{eqnarray}
&&\sum_{s}\bar{v}^s\gamma_5(k\ssh+m)\sigma^{i+}(k\ssh+q\ssh+m)\gamma_5
\nonumber\\
&&\times(k\ssh+q\ssh-P\ssh_\pi+m)\gamma^+v^s\\
&=&\textmd{Tr}[(P\ssh_\pi-k\ssh-m)(k\ssh-m)\gamma^i\gamma^+(k\ssh+q\ssh-m)\nonumber\\
&&\times(k\ssh+q\ssh-P\ssh_\pi+m)\gamma^+]\\
&=&8(P_\pi^+)^2m\mathbf{q}_\p^i
~~~~~~~~~~~~~~~~\textmd{when}~~q^+=0.
\end{eqnarray}
 After integrating over
$\mathbf{q}_\p$, we obtain the expression of
$h_{1\pi}^\p(x,\kp^2)$ in the antiquark spectator model:
\begin{eqnarray}
h_{1\pi}^\p(x,\kp^2)&=&-\frac{g_\pi^2}{2(2\pi)^3}\frac{e_1e_2}{4\pi}\frac{mM_\pi(1-x)}{\kp^2(\kp^2+B_\pi)}
\textmd{ln}\left (\frac{\kp^2+B_\pi}{B_\pi}\right )\nonumber\\
&=&\frac{A_{\pi}}{\kp^2(\kp^2+B_\pi)}\textmd{ln}\left
(\frac{\kp^2+B_\pi}{B_\pi}\right ) \label{h1p},
\end{eqnarray}
with $B_{\pi}$ given in Eq.~(\ref{bpi}) and
\begin{equation}
A_\pi=\frac{g_\pi^2}{2(2\pi)^3}\left
(-\frac{e_1e_2}{4\pi}\right)mM_\pi(1-x).
\end{equation}
In Ref.~\cite{bbh03} $h_1^\perp(x,\kp^2)$ of the proton is
computed in the quark-scalar-diquark model as:
\begin{equation}
h_{1}^\p(x,\kp^2)=\frac{A_p}{\kp^2(\kp^2+B_p)}\textmd{ln}\left
(\frac{\kp^2+B_p}{B_p}\right ).\label{h1}
\end{equation}
Comparing Eq.~(\ref{h1p}) with Eq.~(\ref{h1}) one can find that
the calculated $h_{1\pi}^\p(x,\kp^2)$ in the spectator antiquark
model has a same transverse momentum dependence of
$h_{1}^\p(x,\kp^2)$ of the proton obtained from the
quark-scalar-diquark model, although the $x$ dependence is
different in $A_{p/\pi}$ and $B_{p/\pi}$ respectively, due to the
different mass parameters in them. This feature may not be held
generally, but one can expect that there is close relation between
$h_1^\perp$ distributions of different hadrons since both
functions are generated by the same underling mechanism.

We can also calculate the T-odd distribution of the transversity
distribution of the valence antiquark $\bar{h}_{1\pi}^\p$ of the
pion from the same model, by replacing the antiquark spectator
with the quark spectator. A similar calculation yields
$\bar{h}_{1\pi}^\p=h_{1\pi}^\p$, showing that the magnitudes of
the distributions for two valence quarks are the same.

\begin{figure*}
\begin{center}
\scalebox{0.65}{\includegraphics{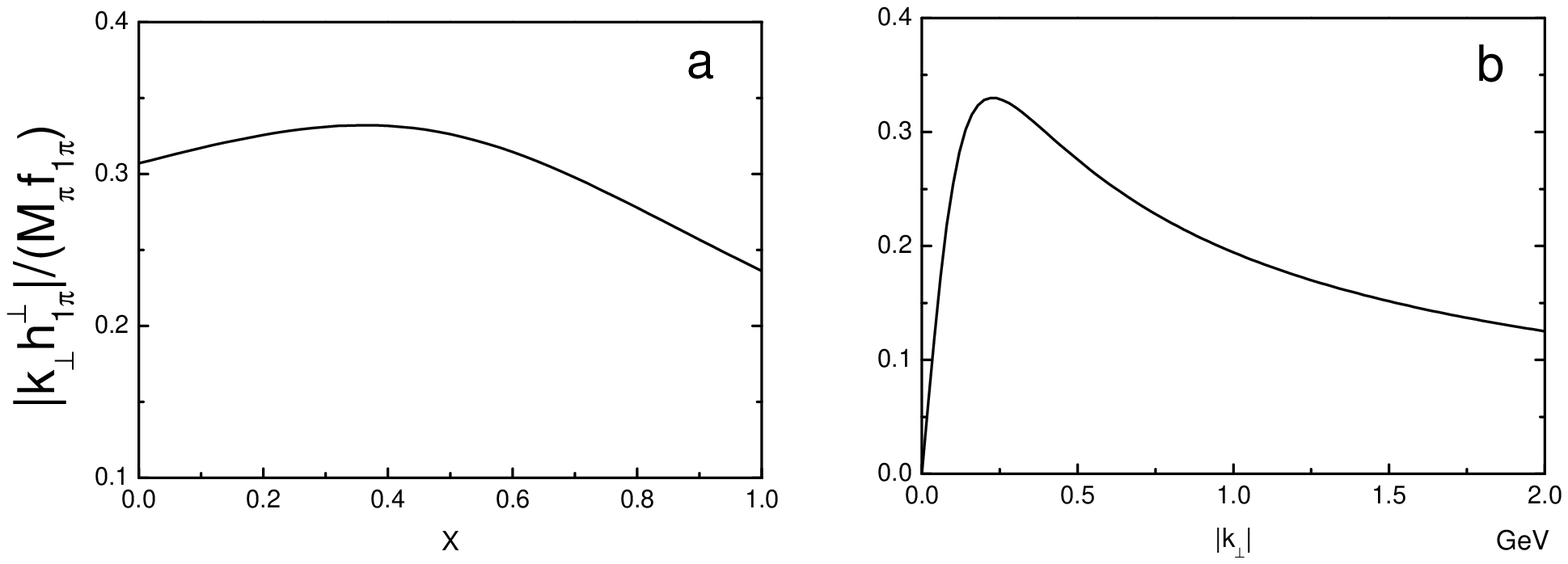}}\caption{\small Model
prediction of $|\kp h_{1\pi}^\p(x,\kp^2)|/[M_\pi
f_{1\pi}(x,\kp^2)]$
 as a function of $x$ and $|\kp|$.}\label{ratio}
\end{center}
\end{figure*}

With $f_{1\pi}(x,\kp^2)$ and $h_{1\pi}^\p(x,\kp^2)$ we estimate
$|\kp h_{1\pi}^\p(x,\kp^2)|/[M_\pi f_{1\pi}(x,\kp^2)]$, which is
proportional to the cos2$\phi$ asymmetries in unpolarized
Drell-Yan process. We choose the mass parameters $m=0.1$~GeV for
the constituent quark mass and $M_\pi=0.137$~GeV for the pion
mass. To generalize our model result to the consequence in QCD we
extrapolate the coupling constant $e_1e_2/4\pi\rightarrow
C_F\alpha_s$, and take $\alpha_s=0.3$ which is adopted in
Ref.~\cite{bhs02a}. We plot the $x$ dependence of the ratio at
$|\kp|=0.3$~GeV in Fig.~\ref{ratio}a and the $|\kp|$ dependence at
$x=0.15$ in Fig.~\ref{ratio}b. The ratios are comparable in
magnitude with $|\kp h_{1}^\p(x,\kp^2)|/[M f_{1}(x,\kp^2)]$ of the
proton (one can see Ref.~\cite{bhs02a}, where $|\kp
f_{1T}^\perp|/(M f_1)$ is given, since $h_1^\perp=f_{1T}^\perp$)
in the quark-scalar-diquark model, where $M$ is the proton mass
and $f_{1}(x,\kp^2)$ is the unpolarized quark distribution of the
proton.

\section{Cos2$\phi$ asymmetries in unpolarized Drell-Yan process}
\begin{figure}
\begin{center}
\scalebox{1}{\includegraphics{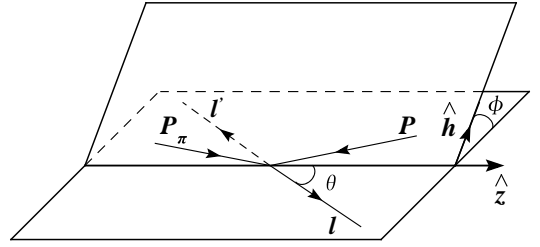}}\caption{\small  Angular
definitions of unpolarized Drell-Yan process in the lepton pair
center of mass frame.}\label{drell-yan}
\end{center}
\end{figure}

The general form of the angular differential cross section for
unpolarized $\pi^- p$ Drell-Yan process is
\begin{eqnarray}
\frac{1}{\sigma}\frac{d\sigma}{\Omega}&=&\frac{3}{4\pi}\frac{1}{\lambda+3}
\left
(1+\lambda\textmd{cos}^2\theta+\mu\textmd{sin}^2\theta\textmd{cos}\phi\right .\nonumber\\
&&\left .+\frac{\nu}{2}\textmd{sin}^2\theta\textmd{cos}2\phi\right
),
\end{eqnarray}
where $\phi$ is the angle between the lepton plane and the plane
of the incident hadrons in the lepton pair center of mass frame
(see Fig.~\ref{drell-yan}). The experimental data show large value
of $\nu$ near to 30\%, which can not be explained by perturbative
QCD. Many theoretical approaches have been proposed to interpret
the experimental data, such as high-twist
effect~\cite{bbkd94,ehvv94}, and factorization breaking
mechanism~\cite{bnm93}. In Ref.~\cite{boer99} Boer demonstrated
that unsuppressed cos$2\phi$ asymmetries can arise from a product
of two chiral-odd $h_1^\perp$ which depends on transverse
momentum. In Ref.~\cite{bbh03} the cos2$\phi$ asymmetry in
unpolarized $p\bar{p}\rightarrow l\bar{l}X$ Drell-Yan process has
been estimated from $h_1^\perp(x,\kp^2)$ for the proton computed
by quark-scalar-diquark model. The maximum of $\nu$ in that case
is in the order of 30$\%$.

In this section we give a simple estimate of cos2$\phi$ asymmetry
in unpolarized $\pi^-p$ Drell-Yan process, from $h_{1\pi}^\perp$
computed by our model. The leading order unpolarized Drell-Yan
cross section expressed in the Collins-Soper frame~\cite{cs77} is
\begin{eqnarray}
&&\frac{d\sigma(h_1h_2\rightarrow l\bar{l}X)}{d\Omega
dx_1dx_2d^2\mathbf{q}_\perp}=
\frac{\alpha^2}{3Q^2}\sum_{a,\bar{a}}\Bigg{\{}
A(y)\mathcal{F}[f_1\bar{f_1}] +B(y)\nonumber\\
&&\times\textmd{cos}2\phi\mathcal{F}\left [(2\hat{\mathbf{h}}\cdot
\mathbf{p}_\perp\hat{\mathbf{h}}\cdot \mathbf{k}_\perp)
-(\mathbf{p}_\perp\cdot
\kp)\frac{h_1^\perp\bar{h}_1^\perp}{M_1M_2}\right
]\Bigg{\}},\nonumber\\
\label{cs}
\end{eqnarray}
where $Q^2=q^2$ is the invariance mass square of the lepton pair,
and the vector $\hat{\mathbf{h}}=\mathbf{q}_\perp/Q_T$. We have
used the notation
\begin{eqnarray}
\mathcal{F}[f_1\bar{f}_1]&=&\int d^2\mathbf{p}_\perp
d^2\kp\delta^2(\mathbf{p}_\perp+\kp-\mathbf{q}_\perp)
f_1^a(x,\mathbf{p}_\perp^2)\nonumber\\
&&\times\bar{f}_1^a(\bar{x},\kp^2).
\end{eqnarray}

From Eq.~\ref{cs} one can give the expression for the asymmetry
coefficient $\nu$~\cite{boer99}:
\begin{eqnarray}
\nu&=&2\sum_{a,\hat{a}}e_a^2\mathcal{F}\left
[(2\hat{\mathbf{h}}\cdot \mathbf{p}_\perp\hat{\mathbf{h}}\cdot
\mathbf{k}_\perp) -(\mathbf{p}_\perp\cdot
\kp)\frac{h_1^\perp\bar{h}_1^\perp}{M_1M_2}\right
]\Bigg{/}\nonumber\\
&&\sum_{a,\bar{a}}e_a^2\mathcal{F}[f\bar{f}]\nonumber\\
&=&\frac{2}{M_1M_2}\frac{\sum_{a,\bar{a}}e_a^2F_a}{\sum_{a,\bar{a}}e_a^2G_a}.
\label{nu}
\end{eqnarray}

\begin{figure}
\begin{center}
\scalebox{0.7}{\includegraphics{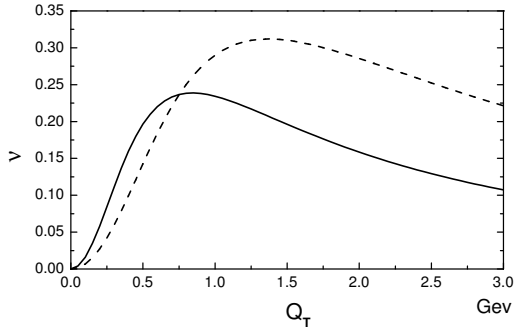}}\caption{\small
Numerical result for the cos2$\phi$ asymmetry in $\pi^-p$
unpolarized Drell-Yan (solid line), using $M_\pi=0.137$ GeV,
m=0.1~GeV, and $x=\bar{x}=0.2$. The dashed line corresponds to the
same asymmetry in $\bar{p}p$ Drell-Yan process estimated in the
quark-scalar-diquark model (see
Ref.~\cite{bbh03}).}\label{cos2phi}
\end{center}
\end{figure}

Our model calculation has shown
$\bar{h}_{1\pi}^\perp=h_{1\pi}^\perp$. Thus in $\pi^-p$
unpolarized Drell-Yan process we can assume $u$-quark dominance,
which means the main contribution to asymmetry comes from
$\bar{h}_{1\pi}^{\perp,\bar{u}}(\bar{x},\kp^2)\times
h_1^{\perp,u}(x,\mathbf{p}_\perp^2)$, since $\bar{u}$ in $\pi^-$
and $u$ in proton are both valence quarks. Then we have $\nu
\approx 2F_u/(M_\pi MG_u)$. To evaluate $\nu$, we use our model
result for $\bar{h}_{1\pi}^{\perp,\bar{u}}$ and $\bar{f}_{1\pi}$,
and we adopt $h_1^{\perp,u}$ and $f_1$ from Ref.~\cite{bbh03}.
Using the $\mathbf{p}_\perp$ integration to eliminate the delta
function in the denominator and numerator in Eq.~(\ref{nu}) one
arrives at
\begin{eqnarray}
F_u&=&\int_0^\infty d|\kp|\int_0^{2\pi}
d\theta_{qk}\frac{A_{\pi}A_p |\kp|}
{\kp^2(\kp^2+B_\pi)}\textmd{ln}\left
(\frac{\kp^2+B_\pi}{B_\pi}\right
)\nonumber\\
&&\times\frac{(\kp^2-2\kp^2\textmd{cos}^2\theta_{qk}+|\kp|
|\mathbf{q}_\perp|\textmd{cos}\theta_{qk})}{(\kp^2+f)(\kp^2+f+B_p)}\nonumber\\
&&\times\textmd{ln}\left (\frac{\kp^2+f+B_p}{B_p}\right ),\label{fu}\\
G_u&=&\int_0^\infty d|\kp|\int_0^{2\pi}
d\theta_{qk}\frac{C_{\pi}C_{p}|\kp|(\kp^2+D_{\pi})}{(\kp^2+B_{\pi})^2}\nonumber\\
&&\times\frac{(\kp^2+f+D_{p})}{(\kp^2+f+B_{p})^2},\label{gu}
\end{eqnarray}
where
$f=\mathbf{q}_\perp^2-2|\mathbf{q}_\perp||\kp|\textmd{cos}\theta_{qk}$,
and $\theta_{qk}$ is the angle between $\kp$  and
$\mathbf{q}_\perp$. In above equations we change the integration
variables into polar coordinates. For the parameters in
Eq.~{\ref{fu}} and Eq.~{\ref{gu}}, we take $A_p/C_p=0.3$ (GeV$^2$)
and $D_p=4B_p\approx1/4$ which are used in Ref.~\cite{bbh03}, and
$A_\pi/C_\pi=0.01$ (GeV$^2$), $D_\pi=2B_\pi\approx0.014$ which
corresponds $\bar{x}=0.2$. We give the numerical evaluation of the
asymmetry as a function of $Q_T$ in Fig.~\ref{cos2phi}, which
shows an unsuppressed result of the order of $\mathcal{O}$ (10\%).
Besides this, the shape of the asymmetry is similar to the
cos2$\phi$ asymmetry in $\bar{p}p$ unpolarized Drell-Yan process
estimated in the quark-scalar-diquark model~\cite{bbh03}.

\section{Conclusion}
In this paper we calculate the (na\"{\i}ve) T-odd $\kp$-dependent
quark transversity distribution $h_1^\perp(x,\kp^2)$ of the pion
for the first time in a quark-spectator-antiquark model in
presence of final-state interaction, and investigate the
cos2$\phi$ asymmetry in unpolarized Drell-Yan process with our
model result. The calculated $h_{1\pi}^\perp(x,\kp^2)$ shows a
same form of transverse momentum dependence as that of
$h_1^\perp(x,\kp^2)$ for the proton computed from a
quark-scalar-diquark model. The similarity of $h_1^\perp(x,\kp^2)$
for different hadrons (for example, nucleon and pseudoscalar
meson) implies that these functions are closely related, because
the mechanism that generates them is the same. Besides this,
$h_{1\pi}^\perp(x,\kp^2)$ is an interesting observable that can
account for the large cos2$\phi$ asymmetry in $\pi^-N$ unpolarized
Drell-Yan process. The contributed asymmetry of
$h_{1\pi}^\perp(x,\kp^2)$ is proportional to $|\kp
h_{1\pi}^\p|/(M_\pi f_{1\pi})$, which is comparable with $|\kp
h_{1}^\p|/(M f_{1})$ of the proton in magnitude. With nonzero
$h_{1\pi}^\perp(x,\kp^2)$ we reveal an unsuppressed cos2$\phi$
asymmetry in unpolarized $\pi^- p$ Drell-Yan process from our
calculation. The shape of the asymmetry is similar to that of
cos$2\phi$ asymmetries in $p\bar{p}$ Drell-Yan process estimated
in Ref.~\cite{bbh03}, suggesting a connection between cos2$\phi$
asymmetries in Drell-Yan processes with different initial hadrons.

\begin{acknowledgments}
This work is partially supported by National Natural Science
Foundation of China under Grant Numbers 10025523 and 90103007.
\end{acknowledgments}

\end{document}